\documentclass{elsarticle}

\begin{document}

\begin{frontmatter}

\title{Nanopore occlusion:
A biophysical mechanism for bipolar cancellation in cell membranes}

\author[MITaddress]{Thiruvallur R. Gowrishankar}
\author[MITaddress]{Julie V. Stern}
\author[MITaddress]{\\Kyle C. Smith}
\author[MITaddress]{James C. Weaver\corref{mycorrespondingauthor}}

\cortext[mycorrespondingauthor]{Corresponding author}
\ead{jcw@mit.edu}

\address[MITaddress]{Harvard-MIT Division of Health Sciences and Technology, Massachusetts Institute of Technology, Cambridge, MA, USA}

\begin{abstract}
Extraordinarily large but short electric field pulses are reported by many experiments to cause
bipolar cancellation (BPC). This unusual cell response occurs if a first pulse is followed by a
second pulse with opposite polarity. Possibly universal, BPC presently lacks a mechanistic
explanation. Multiple versions of the ``standard
model" of cell electroporation (EP) fail to account for BPC. Here we show, for the first time, how an extension of
the standard model can account for a key experimental observation that essentially defines BPC:
the amount of a tracer that enters a cell, and how tracer influx can be decreased
by the second part of a bipolar pulse. The extended model can also account for the recovery of
BPC wherein the extent of BPC is diminished if the spacing between the first and second pulses is increased. Our approach is reverse engineering, meaning that we identify
and introduce an additional biophysical mechanism that allows pore transport to change. We hypothesize that occluding molecules from outside the membrane 
enter or relocate within a pore. Significantly, the additional mechanism
is fundamental and general, involving a combination of partitioning and hindrance. Molecules near the membrane can enter pores to block transport of tracer molecules while still
passing small ions (charge number $\mathrm{\pm 1}$) that govern electrical behavior. Accounting for such behavior requires an extension of the standard model.
\end{abstract}


\end{frontmatter}


\section{Introduction}

Over the past several years several publications have reported and partially characterized the phenomenon of ``bipolar cancellation" (BPC), using a variety of in vitro experiments with isolated
cells \cite{IbeyEtAlPakomov_Cancellation_BBRC2013,%
PakhomovEtAlIbey_CancellationCellularResponses-nsPEF-StimulusPolarityReversal_CellMolLifeSci2014,
GianulisEtAlPakhomov_ElectroporationMammalianCellsNanosecondFieldOscillations-FieldReversalInihibition_SciReports2015,%
MerlaEtal_FreqSpectrum_Um_BPC_BBA_2017,%
ValdezEtal_AsymmBipolarNspef_pulsewidth_BPC_SciRep_2017,%
PakhomovEtal_SecondPhase_BPC_EP_Efficiency_Bioelectrochem2018,%
Ibey_BPCBookChapter_Miklavcic2018,%
GianulusEtAlPakhomov_ElectropermeabilizationUniorBipolar_nsPEF-ImpaceOfExtracellularConductivity_Bchem_2018%
}. 
BPC manifests as reduction or cancellation of bioeffects, specifically the uptake of tracers such as YO-PRO-1, propidium or calcium.  BPC occurs when two pulses of opposite polarity (not necessarily of same amplitude) are applied in rapid succession
\cite{IbeyEtAlPakomov_Cancellation_BBRC2013,%
PakhomovEtAlIbey_CancellationCellularResponses-nsPEF-StimulusPolarityReversal_CellMolLifeSci2014,
GianulisEtAlPakhomov_ElectroporationMammalianCellsNanosecondFieldOscillations-FieldReversalInihibition_SciReports2015,%
MerlaEtal_FreqSpectrum_Um_BPC_BBA_2017,%
ValdezEtal_AsymmBipolarNspef_pulsewidth_BPC_SciRep_2017,%
PakhomovEtal_SecondPhase_BPC_EP_Efficiency_Bioelectrochem2018,%
Ibey_BPCBookChapter_Miklavcic2018,%
GianulusEtAlPakhomov_ElectropermeabilizationUniorBipolar_nsPEF-ImpaceOfExtracellularConductivity_Bchem_2018%
}.   The extent of cancellation decreases with increased separation of the two opposite polarity pulses.  

One striking feature is that BPC requires short, very large fields (nsPEF or nanosecond pulsed
electric fields).  These are not the longer, smaller field pulses used in conventional cell electroporation (EP)
since the 1970s  
\cite{WeaverEporeReviewJCellularBiochem1993,%
Weaver_Review_ElectroMed99SpecialIssue_IEEE_TransactionsPlasmaScience2000,%
WeaverEtAl_BriefOverviewStrength-Duration-%
AdditionalIntracellularEffects_Bchem2012%
}, but are nsPEF pulses used in supra-EP studies \cite{BeebeEtAlSchoenbach_NanosecondPulseApoptosisTumorInhibition_IEEE_TransPlasmsSci2002,%
SchoenbachEtAl_UltrashortPulsesNewGatewayReview_ProcIEEE2004,%
Frey_PlasmaMembraneVoltageChangesDuringNanosecondPulsedElectricFields_BPJ2006,%
NuccitelliEtal_nsPEFSelfDestructMelanomas_BBRC2006NuccitelliEtal_nsPEFSelfDestructMelanomas_BBRC2006,%
SchoenbachEtAl_BioelectricEffectsIntenseNanosecondPulsesReview_IEEE_TransDieIns2007%
}.
While not yet understood mechanistically, BPC is reported to mainly occur for widely separated mammalian cells in
vitro, for applied electric field pulse strengths of 4 - 100 kV/cm and durations of 10 to
600 ns. 

In some experiments pulse trains predominate, which greatly complicates interpretation because of
memory effects due to pore lifetimes.  
Other studies employ single pulses, which is more relevant to basic
understanding, and therefore the focus of the present work
\cite{KrassowskaFilev_ModelingElectroporationSingleCellSphericalPoreExpansion_BPJ2006,%
SonEtAl_BasicFeaturesCellModelEP-TwoDifferentPulses-CalceinPropdium_22JUL2014Published_Corr09JUL2015_JMB2014}.

An unusual BPC feature is that the second part (reversed polarity) of the pulse should move tracer molecules ``uphill", against the concentration gradient. The potential implications of BPC are tantalizing, but initial explanatory hypotheses have failed.
To our knowledge, the present paper is the first report of a biophysical model that
can account for functional features of BPC. 

Significantly, attempts to use the standard cell EP
model to account for BPC all failed. The standard model always predicts a large number of pores such that the diffusive influx always leads to an increase in intracellular tracer molecule.  Essentially all EP delivery/extraction protocols accelerate transport down a solute concentration gradient.  For BPC the second pulse should do the opposite.  This apparently simple change greatly increases the problem difficulty: how can tracer molecule entry be slowed?

The standard EP model is based on lipidic transient pores
(TPs) that form in lipid bilayer membranes in contact with aqueous electrolytes on both sides,
and is consistent with many experimental observations
\cite{KrassowskaFilev_ModelingElectroporationSingleCellSphericalPoreExpansion_BPJ2006,%
LiLin_NumericalSimulationMolecularUptakeElectroporation_BChem2011,%
SmithKC_CellModelWithDyamicPoresAndElectrodiffusionofChargedSpecies_DoctorateThesisMIT_2011,%
SonEtAl_PulseTrainConventionalEP-20FoldDecreaseThenSawtooth-CalciumUpake-Electrosensitization_TBME2015}.
The standard model is essentially an
extension of the Schwann model for either spherical or cylindrical cells
\cite{KrassowskaFilev_ModelingElectroporationSingleCellSphericalPoreExpansion_BPJ2006,%
SmithKC_CellModelWithDyamicPoresAndElectrodiffusionofChargedSpecies_DoctorateThesisMIT_2011,%
SmithKC_CellModelWithDyamicPoresAndElectrodiffusionofChargedSpecies_DoctorateThesisMIT_2011}.  
By adding TP creation for
supra-physiologic transmembrane voltages the resulting model exhibits non-linear TP creation that
begins at one (anodic) pole, followed by poration at the other (cathodic) pole, and then with time
during a porating pulse, additional pore creation further away from the cell's poles 
\cite{KrassowskaFilev_ModelingElectroporationSingleCellSphericalPoreExpansion_BPJ2006,GowrishankarEtAl_DistributionOfFieldsPotentialsElectroporationSites_ConventionalSupraEP_CellModelExplicitOrganelles_BBRC2006}.

\begin{figure*}
\begin{tabular}{lll}
{\large\textbf A} \hspace*{1.3in}& 
{\large\textbf B} \hspace*{1.3in}& 
{\large\textbf C} \\
\includegraphics[width=1.4in]{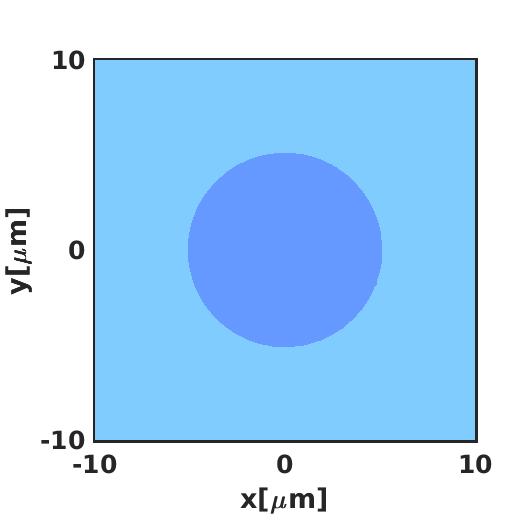} &
\includegraphics[width=1.4in]{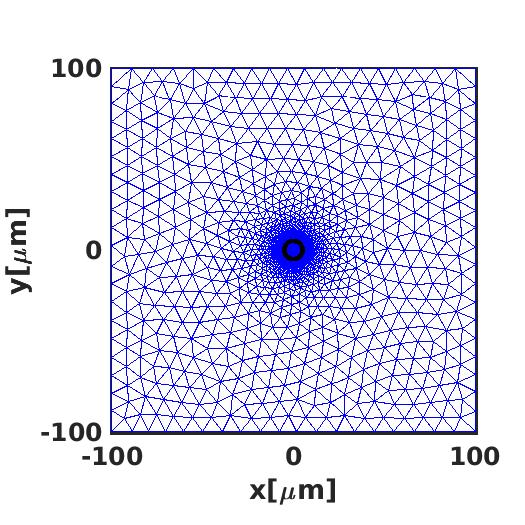} &
\includegraphics[width=1.4in]{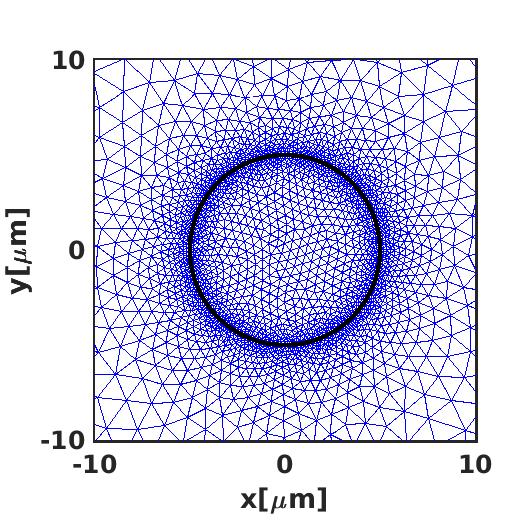}\\
\end{tabular}
\caption{\textbf{Isolated cell model.} The 5 $\mathrm{\mu m}$ radius cylindrical cell (A) is contained in a $\mathrm{200\ \mu m\times 200 \ \mu m}$ system model (B).  The cell mesh transport network model (MTNM)  \cite{SmithKC_CellModelWithDyamicPoresAndElectrodiffusionofChargedSpecies_DoctorateThesisMIT_2011} is represented by 150 transmembrane node-pairs (C) that describe local transmembrane voltages, pore distributions, hindrance, partitioning of molecules and ions into the pores, and molecular transport.  The 4-nm thick membrane has a resting potential of -50 mV due to a fixed current source \cite{SmithKC_CellModelWithDyamicPoresAndElectrodiffusionofChargedSpecies_DoctorateThesisMIT_2011}.  The field is created by applying external pulse generator voltages to the top 
and bottom rows of nodes of the simulation box.  Each of the local areas associated with a transmembrane node-pair is regarded as a very small planar membrane patch (a Voronoi cell) endowed with a resting potential source and a complete dynamic EP model \cite{SmithKC_CellModelWithDyamicPoresAndElectrodiffusionofChargedSpecies_DoctorateThesisMIT_2011}.}
\label{fig_1}
\end{figure*}

Here we propose a mechanism for BPC that is based on increased pore occlusion and a corresponding decrease in tracer transport due to the presence of charged molecules within pores.  The increased occlusion is represented in the model by a decrease in the occlusion factor magnitude.  We also account for the possibility of a weak interaction of the inserted molecule with the membrane pore by allowing the recovery of the occlusion factor.

\section{Methods}

\subsection{Cell Electroporation EP model}

Cell EP inescapably involves spatially distributed, highly nonlinear and
hysteretic  interactions  throughout  a  cell system  model.  We use a cylindrical cell membrane contacting electrically conducting extracellular and intracellular media
\cite{SmithWeaver_ActiveMechanismsNeededSubmicrosecondMVperMeterPulses_BPJ2008,%
SmithKC_CellModelWithDyamicPoresAndElectrodiffusionofChargedSpecies_DoctorateThesisMIT_2011,%
GowrishankarEtAl_TransportBasedModelsElectricFieldCellResponses_Review_ProcIEEE2012,%
SonEtAl_BasicFeaturesCellModelEP-TwoDifferentPulses-%
CalceinPropdium_22JUL2014Published_Corr09JUL2015_JMB2014,%
SonEtAl_PulseTrainConventionalEP-20FoldDecreaseThenSawtooth-CalciumUpake-%
Electrosensitization_TBME2015}.
These complex interactions are solved computationally with an isolated cylindrical cell model (Fig.~\ref{fig_1}).  We describe the system using the meshed transport network model (MTNM) elsewhere (above publications).
The cylindrical plasma membrane (PM) has $5 \mathrm{\, \mu m}$ radius, $6.7 \mathrm{\, \mu m}$ height, and 4~nm thickness (Fig.~1A).
The extracellular region is represented by 2077 nodes (or Voronoi cells. which are the local regions),
and the intracellular region is represented by 891 nodes.  Of these, 150
node pairs (one extracellular and one intracellular) span the PM (Fig.~\ref{fig_1}B).
The two electrolytes are represented by passive models that describe charge transport and storage within the electrolyte
\cite{GowrishankarWeaver_TransportLatticeSimulation_PNAS2003}.  The PM node pairs (Fig.~\ref{fig_1}C) contain a complete dynamic EP model that provides the local kinetics of membrane pore creation, evolution, and destruction, and include associated changes in transmembrane voltage and membrane conductance.  

We use $D_{\mathrm{p}}$ = $2 \times 10^{-13} \mathrm{\, m^2/s}$,
for the diffusion coefficient in pore radius space and a maximum pore radius, $r_{\mathrm{p,max}}$ of 12~nm with a pore lifetime of 100~s.  The details of the local membrane EP model are described elsewhere
\cite{SmithWeaver_ActiveMechanismsNeededSubmicrosecondMVperMeterPulses_BPJ2008,
SmithKC_CellModelWithDyamicPoresAndElectrodiffusionofChargedSpecies_DoctorateThesisMIT_2011,%
GowrishankarEtAl_TransportBasedModelsElectricFieldCellResponses_Review_ProcIEEE2012,
SonEtAl_BasicFeaturesCellModelEP-TwoDifferentPulses-CalceinPropdium_22JUL2014Published_Corr09JUL2015_JMB2014}.
The local membrane models also include a -50 mV resting transmembrane voltage source.  
Other parameters for describing membrane EP within local membrane areas
(regions associated with a transmembrane node pair)
and adjacent aqueous media are given elsewhere
\cite{VasilkoskiEtAl_ElectroporationAbsouteRateEquationNanosecondPoreCreation_PhysRevE2006}.

\subsection{Occluded transport}

 We assume that once a lipidic transient pore (TP) is created in the membrane, one or more charged molecules enter the pore.  The presence of a charged molecule in the pore causes occlusion that hinders the movement of ions and tracer molecules.  Some of the molecules are weakly bound to the pore wall and with time leave the pore.  However, in the case of a bipolar pulse, the second pulse draws more molecules into the pores, increasing occlusion.   

We modify the standard model of electroporation by introducing an occlusion factor, $\mathrm{O(t)}$, that accounts for a decrease in pore-mediated transport of both small ions and tracer molecules. 
$\mathrm{O(t)}$ represents the total occlusion due to external molecule hindrance and partitioning.  In addition, O(t) kinetics can account for the partial recovery of the membrane by the release of weakly bound molecules from the pore walls.  $\mathrm{O(t)}$ accounts for the decrease in tracer transport through a pore in the presence of external molecules in the pore.

\begin{figure}[!h] 
\centering
\begin{tabular}{ll}
{\large\textbf A (BP 5 peaks)} & {\large\textbf B (UP 3 peaks)}\\
\includegraphics[width=2.2in]{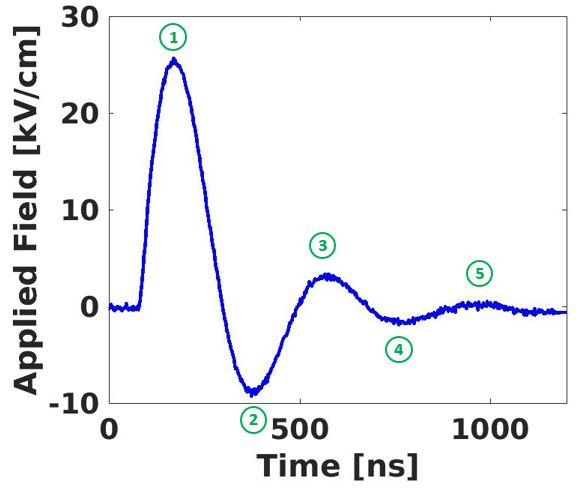}&
\includegraphics[width=2.2in]{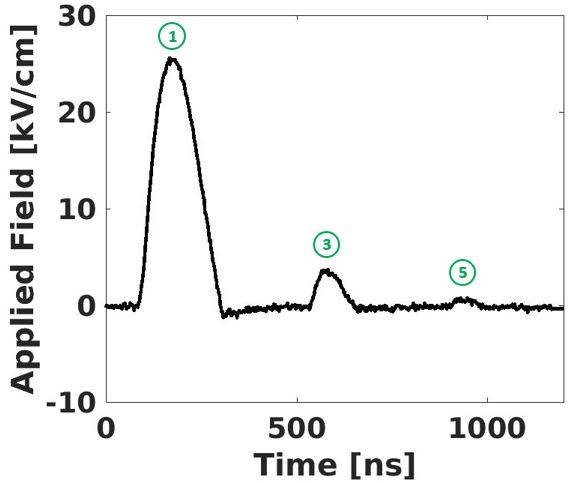}\\
{\large\textbf C} & {\large\textbf D}\\
\includegraphics[width=2.2in]{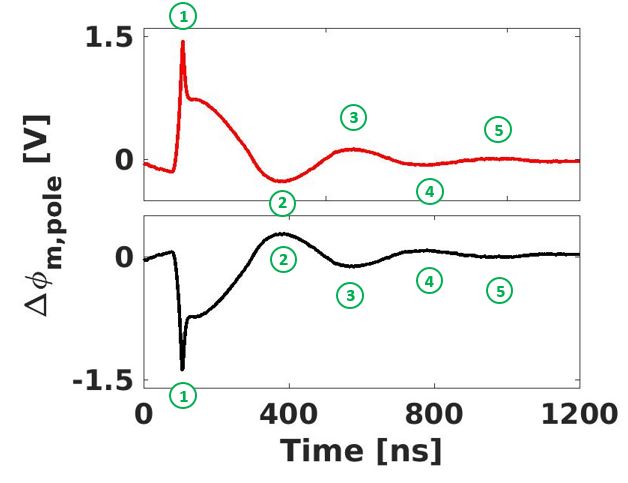}&
\includegraphics[width=2.2in]{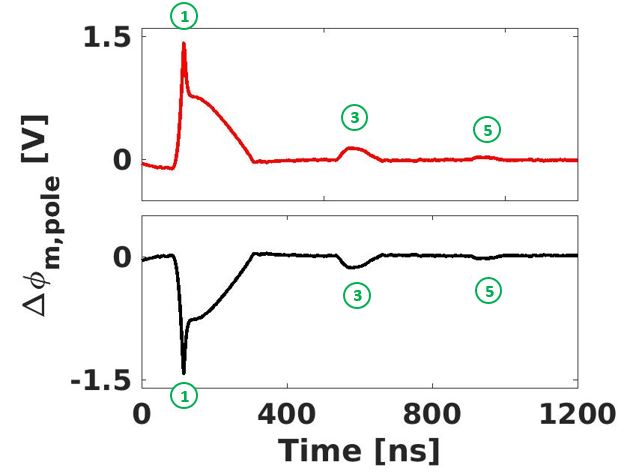}\\
\end{tabular}
\caption{\textbf{Electrical response to bipolar (BP) and unipolar (UP pulses.}  The BP (A) and UP (B) pulses are digitized versions of experimental waveforms
\cite{GianulisEtAlPakhomov_ElectroporationMammalianCellsNanosecondFieldOscillations-FieldReversalInihibition_SciReports2015}.  Each pulse of the complex waveform is approximately 200 ns long and the amplitude of the first positive peak is 24 kV/cm.  The BP has 5 peaks (A) and UP has 3 peaks (B).  Peaks 2 and 4 in the BP are not present in the UP. The UP is derived from the BP by rectification.   voltage, $\mathrm{\Delta \phi_m}$, response is shown at the anodic (red) and cathodic (black)
poles for the BP (C) and UP (D) pulses.  In both unipolar and bipolar cases, $\mathrm{\Delta \phi_m}$ increases rapidly at the onset of the pulse.  This rise initiates a burst of pore creation, increasing the conductance of the membrane.  The conductance increase brings down $\mathrm{\Delta \phi_m}$ to a plateau of 0.7 V before the pulse starts declining.} 
\label{fig_2}
\end{figure}

\subsection{Applied field}

Our model can readily accommodate experimental waveforms with complex characteristics, including a decaying sinusoid.  
We model the response of two different but related electric field pulses: bipolar (BP; + and - 24 kV/cm, 200+200 ns; Fig.~\ref{fig_2}A) and unipolar (UP; 24 kV/cm, 200 ns; Fig.~\ref{fig_2}B).   These pulses are digitized version of the experimental pulses of Gianulis {\it et al.} \cite{GianulisEtAlPakhomov_ElectroporationMammalianCellsNanosecondFieldOscillations-FieldReversalInihibition_SciReports2015}.  

\subsection{Electrolytes}

The extracellular and intracellular media have electrical conductivities of 1.2 $\mathrm{S/m}$ and 0.3 $\mathrm{S/m}$, respectively.  The extracellular medium also contains $\mathrm{1\ \mathrm{\mu}M}$ YO-PRO-1 (YP), a fluorescent dye with molecular properties: charge number: +2, molecular length: 1.7 nm, molecular radius: 0.53 nm, 
extracellular diffusion coefficient: $\mathrm{5.39\times 10^{-10}\  m^2/s}$, and 
intracellular diffusion coefficient: $\mathrm{1.35\times 10^{-10}\  m^2/s}$
\cite{SmithKC_CellModelWithDyamicPoresAndElectrodiffusionofChargedSpecies_DoctorateThesisMIT_2011}.

\section{Results}

The uptake of YP in an isolated cell model is compared to the experimental uptake from UP and BP pulses
\cite{GianulisEtAlPakhomov_ElectroporationMammalianCellsNanosecondFieldOscillations-FieldReversalInihibition_SciReports2015}.  
The effect of changing hindrance due to entry of external molecules is also presented.  Our model's response can therefore account for the reported experimental BPC behavior.

\subsection{Electrical response to a bipolar and unipolar pulse}

Figure~\ref{fig_2}A and \ref{fig_2}B show transmembrane voltage ($\mathrm{\Delta \phi_m}$) at the anodic and cathodic poles of the cell in response to the BP (A) and UP (B) pulses of Fig.~\ref{fig_2}.  $\mathrm{\Delta \phi_m}$ increases with the onset of the applied field until the associated increase in membrane conductance causes the reversible electrical breakdown (REB) of the membrane.  The REB peak occurs within 25 ns from the start of the pulse.   Only the first peak of the pulse causes REB of the membrane.   The magnitudes of the subsequent peaks of the applied field are not large enough to cause REB given persisting conductance.  $\mathrm{\Delta \phi_m}$ responses for BP and UP at the poles are also bipolar and unipolar, respectively.

\subsection{Occlusion}

Figure SI-1 (Supplemental Information) shows the hindrance factor as a function of pore radius for YP.
The hindrance factor, $\mathrm{H}$ ($\mathrm{0 \leq H \leq 1}$) without occluding molecules is determined by the size of YP molecule relative to the pore radius
\cite{SmithKC_CellModelWithDyamicPoresAndElectrodiffusionofChargedSpecies_DoctorateThesisMIT_2011}.  When $\mathrm{H\rightarrow 0}$, transport is significantly hindered, close to zero.  However, when $\mathrm{H\rightarrow 1}$, transport is largely unhindered, which leads to transport rates approaching bulk electrolyte values.  In other words, a higher value of hindrance factor corresponds to less transport.  For YP and pore radii less than 2 nm, YP molecules experience significant hindrance for transport through a minimum-sized (0.85 nm radius) pore.  In this example, transport of YP is reduced by a factor of 0.007 for a pore radius of 0.85 nm and 
by a factor of 0.2 for a pore radius of 2 nm.  However, if the pore is obstructed by external molecules, the increased hindrance extends to larger pores. Both BP and UP pulses considered here cause pores to expand to no more than 2 nm in pore radius (Fig.~SI-2). 

\subsection{Effect of hindrance on YP uptake}

Submicrosecond pulses cause supra-EP (large number of small pores)
\cite{GowrishankarEtAl_DistributionOfFieldsPotentialsElectroporationSites_ConventionalSupraEP_CellModelExplicitOrganelles_BBRC2006}.  Small pores limit the uptake of molecules like YP.  But even pores as small as 2 nm radius allow YP (length: 1.7 nm, radius: 0.53 nm) to cross the membrane with a small hindrance factor.  Both BP and UP create nearly identical distribution of pores (size and number) (Fig SI-2).   However, when H decreases, the uptake decreases.  Both 
BP and UP show similar uptake profiles for different H values 
because of similar pore distribution for both pulses.

\begin{figure*}[!t]
\centering
\begin{tabular}{ll}
{\large\textbf A} & {\large\textbf B}\\
 \includegraphics[width=2.3in]{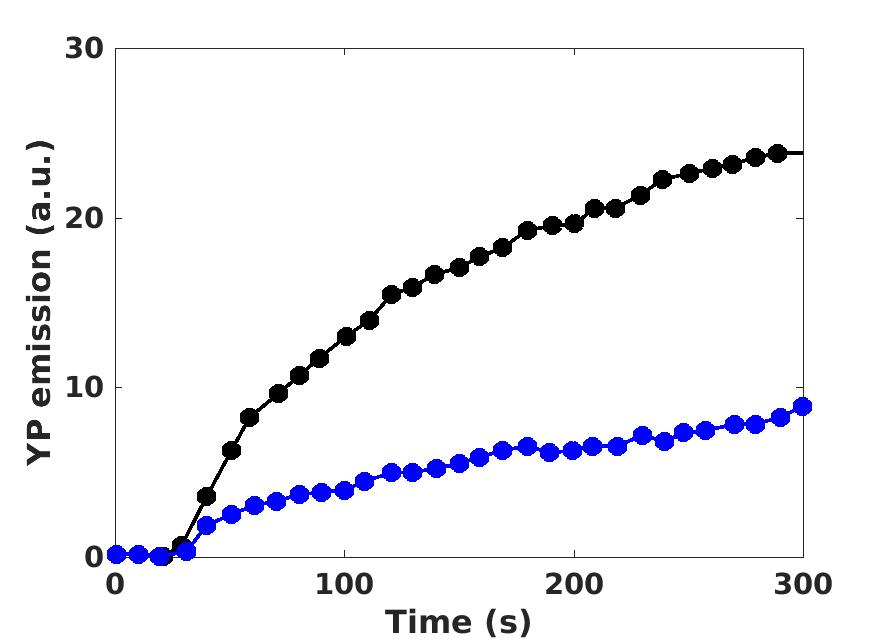}&
\includegraphics[width=2.1in]{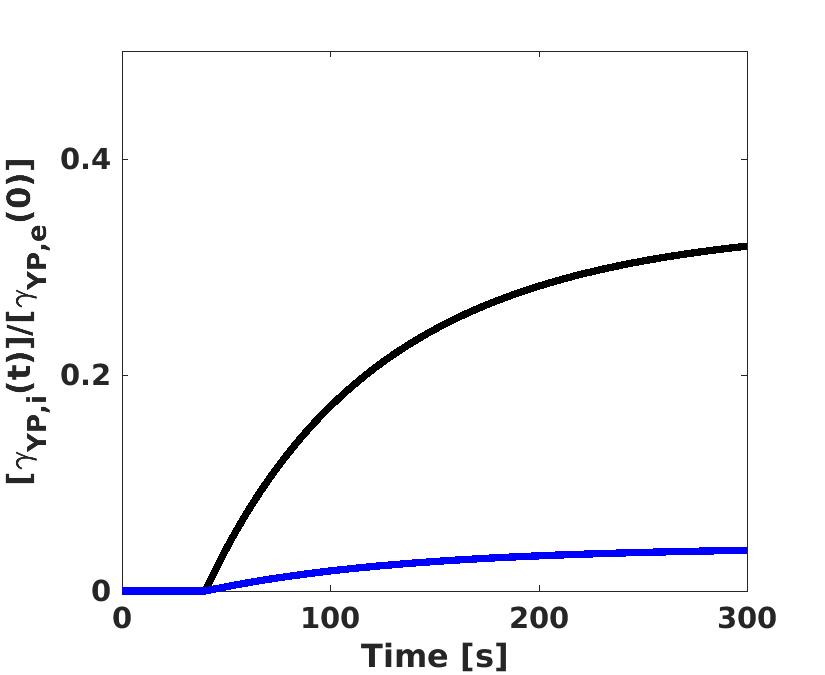}\\
\end{tabular}
\caption{\textbf{Cell model YP uptake compared to experiment\cite{GianulisEtAlPakhomov_ElectroporationMammalianCellsNanosecondFieldOscillations-FieldReversalInihibition_SciReports2015}.} Experimental uptake of YP in CHO cells subject to a BP (blue) and UP (black) pulses are shown (A).  The corresponding uptake in the cell model is shown in (B) normalized to the initial extracellular concentration of YP. Hindrance, quantified by the H-factor (hindrance factor) was increased at 200 ns for UP by 100 times and for BP by 1000 times.  Our model reasonably accounts for bipolar cancellation (decreased uptake for a BP) as seen in experiments\cite{GianulisEtAlPakhomov_ElectroporationMammalianCellsNanosecondFieldOscillations-FieldReversalInihibition_SciReports2015}.} 
\label{fig_3}
\end{figure*}

\subsection{Occlusion reduces molecular uptake for a bipolar pulse}

Gianulis {\it et al.} 
\cite{GianulisEtAlPakhomov_ElectroporationMammalianCellsNanosecondFieldOscillations-FieldReversalInihibition_SciReports2015}
show that bipolar and unipolar nanosecond electric field pulses (Fig.~\ref{fig_2}) enable electroporative uptake of YP in CHO cells (Fig.~\ref{fig_3}A).   
The experimental study shows that uptake from a BP is three times smaller than that from a corresponding UP.

The intracellular YP concentration in the isolated cell model shows that a decreased uptake from a BP (compared to a corresponding unipolar pulse) corresponds to increased occlusion (less uptake).  Figure ~\ref{fig_3}B shows YP uptake for UP ($\mathrm{O(t>}$200 ns) = 0.01) and for bipolar pulse ($\mathrm{O(t>}$200 ns) = 0.001).  The decrease in YP uptake for a BP compared to the UP agrees with experimental results
\cite{GianulisEtAlPakhomov_ElectroporationMammalianCellsNanosecondFieldOscillations-FieldReversalInihibition_SciReports2015}.  However, the uptake ratio (BP:UP) at t = 300~s is nearly twice as large in our model compared to the experimental values.

\begin{figure*}[!t]
\centering
\begin{tabular}{ll}
{\large\textbf A} & {\large\textbf B}\\
\includegraphics[width=2.5in]{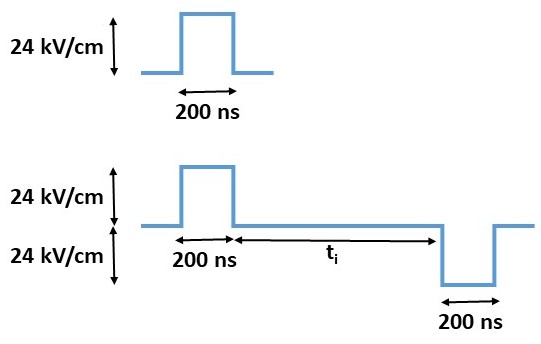}&
\includegraphics[width=2.1in]{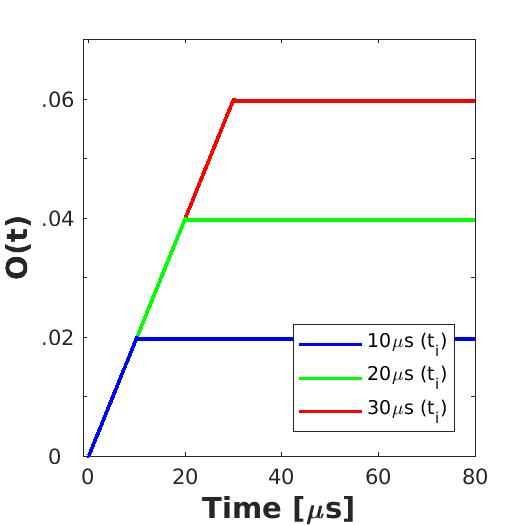}\\
\multicolumn{2}{c}{\large\textbf C}\\
\multicolumn{2}{c}{\includegraphics[width=3.2in]{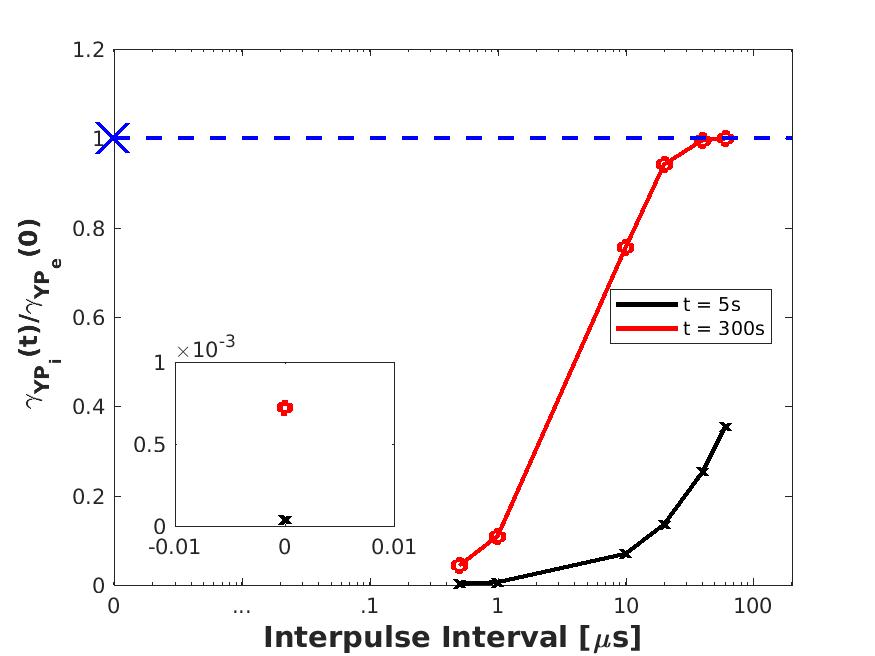}}\\
\end{tabular}
\caption{\textbf{Occlusion recovery in a single cell model.} 
\textbf{(A)} The UP is an idealized trapezoidal pulse (24 kV/cm, 200 ns).  An identical pulse of opposite polarity occurs after an interpulse interval, $\mathrm{t_i}$. 
\textbf{(B)} Occlusion factor, O(t) is scaled linearly increasing from $\mathrm{10^{-5}}$ to $\mathrm{10^{-1}}$ during a $\mathrm{t_i}$ of 0 to 50 $\mathrm{\mu}$s.   As a large number of pores ($\mathrm{10^5}$) is created at the start of the pulse, occluding molecules enter some pores causing significant occlusion.  But after the pulse, loosely bound molecules depart leading to larger O(t) for wider interpulse intervals. \textbf{(C)} Uptake of YP at t = 5 s and t = 300 s for different interpulse intervals, normalized to initial extracellular YP concentration.   Uptake from having no interpulse interval is shown in the inset.  The large X at 0 represents the level of normalized uptake from the UP, since for a UP no interpulse interval applies.  The dotted line gives the maximum relative YP concentration ratio.
} 
\label{fig_4}
\end{figure*}

\subsection{Interpulse duration dependence}

BPC is demonstrated by reduced uptake when a second pulse of opposite polarity follows a first pulse.  However, if the second pulse is applied after a delay, BPC is diminished
\cite{PakhomovEtAlIbey_CancellationCellularResponses-nsPEF-StimulusPolarityReversal_CellMolLifeSci2014}.  We accordingly extend the mechanistic hypothesis of molecular pore occlusion to include a binding strength effect.  Following the first pulse, weakly interacting occluding molecules leave the pore with an assumed linear time dependence over 50 $\mathrm{\mu s}$.  However, if a second pulse of opposite polarity is applied sooner ($\mathrm{t_i<50\ \mu s}$) after the first pulse 
(Fig.~\ref{fig_4} (A)), it slows occluding molecules exiting the pore and thus extends time of occlusion (Fig.~\ref{fig_4} (B)), where YP uptake is shown at 5 s and 300 s after a BP in Fig.~\ref{fig_4} (C).   Uptake from a BP is smaller than UP (denoted by x) on black curve for short interpulse intervals demonstrating BPC.  However, when the interpulse interval is longer than 10 $\mathrm{\mu }$s, BPC  is diminished, with uptake approaching the UP values.  The recovery of BPC effect is faster for fields of larger amplitude (Fig.~SI-3).

\section{Discussion}

Different versions of standard cell EP models all focus on the lipidic portion of the plasma membrane \cite{KrassowskaFilev_ModelingElectroporationSingleCellSphericalPoreExpansion_BPJ2006,%
LiLin_NumericalSimulationMolecularUptakeElectroporation_BChem2011,%
SmithKC_CellModelWithDyamicPoresAndElectrodiffusionofChargedSpecies_DoctorateThesisMIT_2011,%
SonEtAl_PulseTrainConventionalEP-20FoldDecreaseThenSawtooth-CalciumUpake-Electrosensitization_TBME2015}.  However, the standard EP model cannot explain BPC in cells as it does not  involve non-membrane molecules (contaminants \cite{MelikovEtAlChizmadzhev_NonconductivePrePore_MetastableSinglePores_BLM_PBJ2001}, extracellular molecules \cite{LeeEtAlSurfactantInducedSealingEpermSkeletalMuscleMembraneInVivoPNAS1992,%
WeaverEtAlHeparinFootInTheDoorBBRC1997,%
CollinsEtAlLee_StructuralFunctionalRecoverySkeletalMuscleEporePoloxamer188_BBA2007} or intracellular molecules).
The revised EP model takes into account such molecules by considering external sources of occluding molecules.  The model could 
be partially tested by purposefully adding occluding molecule candidates to the  extracellular medium.

\subsection{Pore size distribution and its effect on molecular transport}

When an external electric field is applied to a cell, pore size distributions evolve from a thermalized distribution around 0.85 nm to larger pore radii.  But, the pulses of Fig.~\ref{fig_2}A and \ref{fig_2}B exist only for 200 ns, not long enough to cause significant pore expansion. Given the long pore lifetime ($\mathrm{\tau_p}$ = 100 s), the pore distribution does not change significantly during the pulse.   

Further, rapid creation of nearly $\mathrm{10^6}$ pores causes several orders of magnitude increase in membrane conductance.  This sudden new electrical load (large membrane conductance) holds down $\mathrm{\Delta \phi_m}(t)$ and leads to the creation of only a small number of additional pores with subsequent pulses.   Although the total pore number, $\mathrm{N(t)}$, is large, most pores are less than 1.5 nm in radius.  These small pores offer significant hindrance to the uptake of YP.

\subsection{Mechanism of bipolar cancellation}

Here we propose a mechanism for bipolar cancellation based on pore occlusion due to entry of external molecules into the membrane
\cite{LeeEtAlSurfactantInducedSealingEpermSkeletalMuscleMembraneInVivoPNAS1992,%
WeaverEtAlHeparinFootInTheDoorBBRC1997,%
MelikovEtAlChizmadzhev_NonconductivePrePore_MetastableSinglePores_BLM_PBJ2001,%
CollinsEtAlLee_StructuralFunctionalRecoverySkeletalMuscleEporePoloxamer188_BBA2007}.  
Occlusion can be caused by initial entrance into the pore or movement (relocation) of occluding molecules within an existing nanopore (conformational change).  This is similar to protein-bound ion channel conformational changes.

The magnitude of occlusion depends on the extent of interaction of the occluding molecules with the pores and the tracer molecule. Occluding molecules may enter the pore partially, weakly bind to the membrane or fully inserted into the membrane.  Molecules that are weakly bound may leave the pore quickly.  If a pulse of opposite polarity is applied before the occluding molecules leave the pore, they can be reinserted into the pore, further increasing occlusion.   

We compared our model results with single pulse experimental results of BPC 
\cite{GianulisEtAlPakhomov_ElectroporationMammalianCellsNanosecondFieldOscillations-FieldReversalInihibition_SciReports2015} that show a decreased uptake with a BP (compared to a UP of same amplitude) up to 300 s after the pulse (Fig.~\ref{fig_3}).   The experimental 300 s time-scale is consistent with our model's $\mathrm{\tau_p}$ = 100 s. Our model suggests that occluding molecules entering pores during a UP (24 kV/cm, 200 ns) can hinder transport by a factor of 100 compared to standard pore transport.  In contrast, a BP of equal amplitude (24 kV/cm, 200+200 ns) hinders transport by a further factor of 10 (overall factor of 1,000).  
 
\subsection{Reduction of calcium transport (a small ion tracer)}

Bipolar cancellation effects have been demonstrated experimentally as decreased net uptake of calcium, propidium, and Yo-Pro-1 due to a bipolar pulse compared to a unipolar pulse \cite{IbeyEtAlPakomov_Cancellation_BBRC2013,%
PakhomovEtAlIbey_CancellationCellularResponses-nsPEF-StimulusPolarityReversal_CellMolLifeSci2014,
GianulisEtAlPakhomov_ElectroporationMammalianCellsNanosecondFieldOscillations-FieldReversalInihibition_SciReports2015,%
MerlaEtal_FreqSpectrum_Um_BPC_BBA_2017,%
ValdezEtal_AsymmBipolarNspef_pulsewidth_BPC_SciRep_2017,%
PakhomovEtal_SecondPhase_BPC_EP_Efficiency_Bioelectrochem2018,%
Ibey_BPCBookChapter_Miklavcic2018,%
GianulusEtAlPakhomov_ElectropermeabilizationUniorBipolar_nsPEF-ImpaceOfExtracellularConductivity_Bchem_2018%
}.  Here we concentrate on YP because of comparable single pulse experimental results
\cite{GianulisEtAlPakhomov_ElectroporationMammalianCellsNanosecondFieldOscillations-FieldReversalInihibition_SciReports2015}.  Modeling of intracellular calcium, in contrast, is more complicated.  
The (assumed) negative charge of most occluding molecules should affect the transit of doubly charged calcium ($\mathrm{z_s}$ = +2) through nanopores because of partitioning.  Also, the large nsPEF pulses can cause supra-EP of not only the PM, but also the endoplasmic reticulum (ER) membranes
\cite{GowrishankarEtAl_DistributionOfFieldsPotentialsElectroporationSites_ConventionalSupraEP_CellModelExplicitOrganelles_BBRC2006}, which may release calcium from internal stores, a potentially significant complication.  

\subsection{Future model extension}
BPC is the only EP application that attempts to move a tracer against its concentration gradient. For this reason alone, the standard model has been successful in accelerating a ``downhill'' tracer transport. According to the standard EP model, cellular uptake is largely determined by post-pulse pore distributions because the post-pulse duration (100s of seconds) dominates behavior over the duration of the pulse (200 ns) \cite{SmithWeaver_TransmembraneMolecularTransportDuringAfter_nsPEF_BBRC2011}.  For uptake to persist for 100s of seconds, pores must remain open for an order of 100 s.  If pore life time is much longer than the pulse duration, subsequent pulses (of same or opposite polarity) will not create new pores as the membrane conductance can remain large, and hold down $\mathrm{\Delta \phi_m}$  \cite{SonEtAl_PulseTrainConventionalEP-20FoldDecreaseThenSawtooth-CalciumUpake-Electrosensitization_TBME2015}.  In such cases, post-pulse diffusive uptake is nearly equal for both UP and BP pulses.  Even if a small difference exists, it will only cause a greater uptake for a bipolar pulse.  The new, extended EP model that accounts for occlusion is essential for describing bipolar cancellation.

%

\section*{Acknowledgment}

This work was partially supported by an AFOSR MURI grant FA9550-15-1-0517.  
We thank P. T. Vernier and E. B. S{\"o}zer for many discussions, E. C. Gianulis for waveform data and K. G. Weaver for continued computer support. 

\newpage


\pagebreak

\renewcommand\thefigure{SI-\arabic{figure}}

\setcounter{figure}{0}

\begin{center}
{\large\bf Supplemental Information}
\end{center}

%
\ \\
\noindent
\textbf{Occlusion and Molecular transport through pores}\\
\ \\

\noindent 
\textbf{Occluding molecules decrease tracer influx} \\
\\
%
Our aim is to estimate the diminished influx of tracer molecules that occurs during the second part of a bipolar pulse that causes 
bipolar cancellation (BPC).
Accordingly, we modify a standard model of electroporation 
\cite{SmithKC_CellModelWithDyamicPoresAndElectrodiffusionofChargedSpecies_DoctorateThesisMIT_2011}
by introducing an occlusion factor,
$O(t)$, that accounts for a decrease in pore-mediated transport of ions and tracer molecules.
Consistent with notation in 
\cite{SmithKC_CellModelWithDyamicPoresAndElectrodiffusionofChargedSpecies_DoctorateThesisMIT_2011,%
SmithEtAl_EmergencelargePoreSubpopulationDuringElectroporationPulses2013,%
SonEtAl_BasicFeaturesCellModelEP-TwoDifferentPulses-CalceinPropdium_22JUL2014Published_Corr09JUL2015_JMB2014}, we consider a tracer as a solute, ``s'', but recognize that the concepts of hindrance and partitioning are more general,
applying also to tracers (e.g. YP) that move along an interior pore surface or pore wall.
The tracer flux through a pore,
$J_{\mathrm{s,p}}$,
is related to the flux in bulk electrolyte,
$J_{\mathrm{s}}$ by
\begin{center}
$J_{\mathrm{s,p}}$ $=$ $[O(t) H K]$ $J_{\mathrm{s}}$
\end{center}
where $H$ and $K$ are the hindrance and partition factors of the nanopore without occluding molecules
\cite{SmithKC_CellModelWithDyamicPoresAndElectrodiffusionofChargedSpecies_DoctorateThesisMIT_2011}. 
The occlusion factor, $O(t)$,
represents the total tracer occlusion due to entry and relocation of external occlusion molecules.
These external molecules reside for various times within pores, altering hindrance and partitioning for tracers.
In addition, with a time dependence, $O(t)$ accounts for the
partial recovery of the membrane by the release of weakly bound occluding molecules from within pores.
\vspace{0.12in}
%

\noindent \textbf{Occlusion effect for tracers and for small ions}\\
\\
%
The small, highly mobile ions that dominate electrical behavior are
$\mathrm{Na^{+}}$, 
$\mathrm{Cl^{-}}$ 
and
$\mathrm{K^{+}}$.
Bipolar cancellation experiments emphasize tracer influx, not electrical behavior.
We expect the effects on tracers such as
$\mathrm{YO-PRO-1^{++}}$ (YP),
$\mathrm{propidium^{++}}$ (Pro),
and
$\mathrm{Ca^{++}}$ 
to be larger than for the ubiquitous, small ions with charge number $\pm 1$.
Put simply, these small ions are likely get through restricted (occluded) pathways more readily than the
tracers with charge number $+2$.
For YP and Pro tracers size also favors a larger occlusion effect.

Altered occlusion represents alteration of both hindrance to molecular transport and also alteration of partitioning of different tracer molecules in the pore-occluding molecule complex.

\pagebreak

\begin{figure}[!t]
\centering
\begin{tabular}{c}
\includegraphics[width=3.2in]{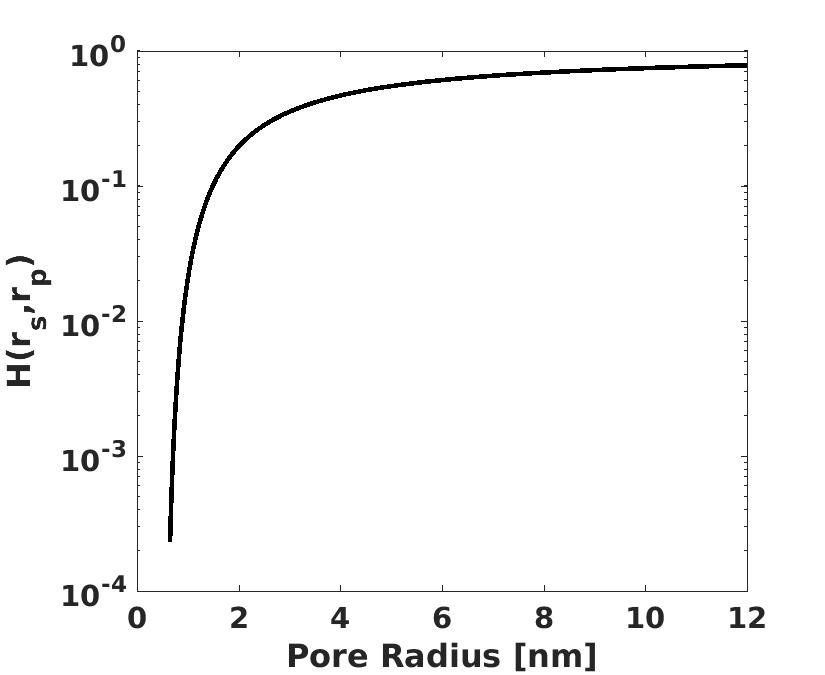}\\
\end{tabular}
\caption{\textbf{Hindrance factor, $\mathbf{H(r_s,r_p)}$, as a function of pore radius, $\mathbf{r_p}$, for YP} which has a radius of 0.53 nm and a length of 1.71 nm.
For pore radii less than 2 nm, YP molecules experience a significant hindrance for transport through the pore.  If the pore is even partially obstructed by an external (non-membrane constituent) molecule, the hindrance extends to larger pores.}  
\label{fig_hindrance}
\end{figure}

\begin{figure*}[!t]
\centering
\begin{tabular}{llll}
\ & 
\textbf{(A)} Before pulse & \textbf{(B)} Peak 1 & \textbf{(C)} Peak 2\\
\ & 
\includegraphics[width=1.25in]{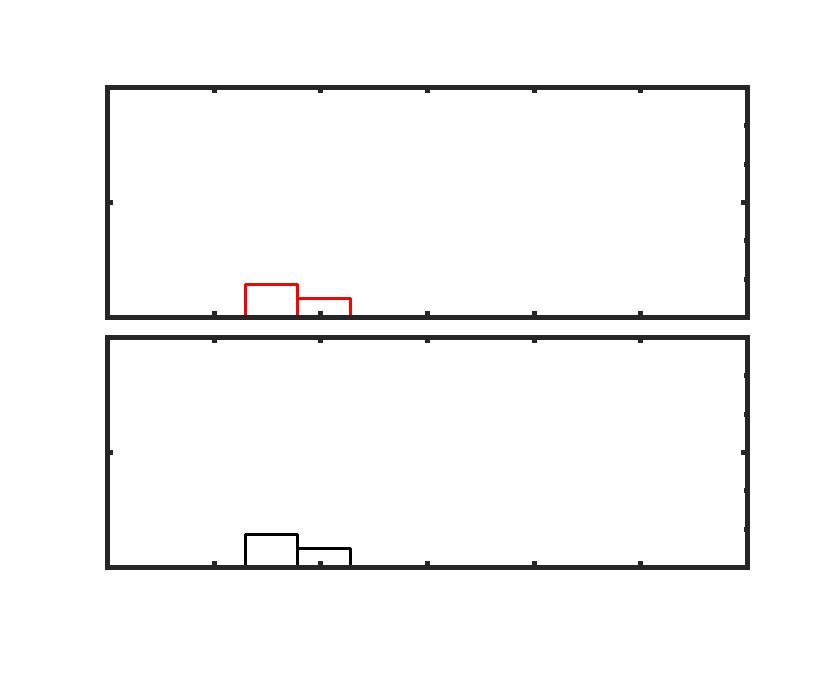}&
\includegraphics[width=1.25in]{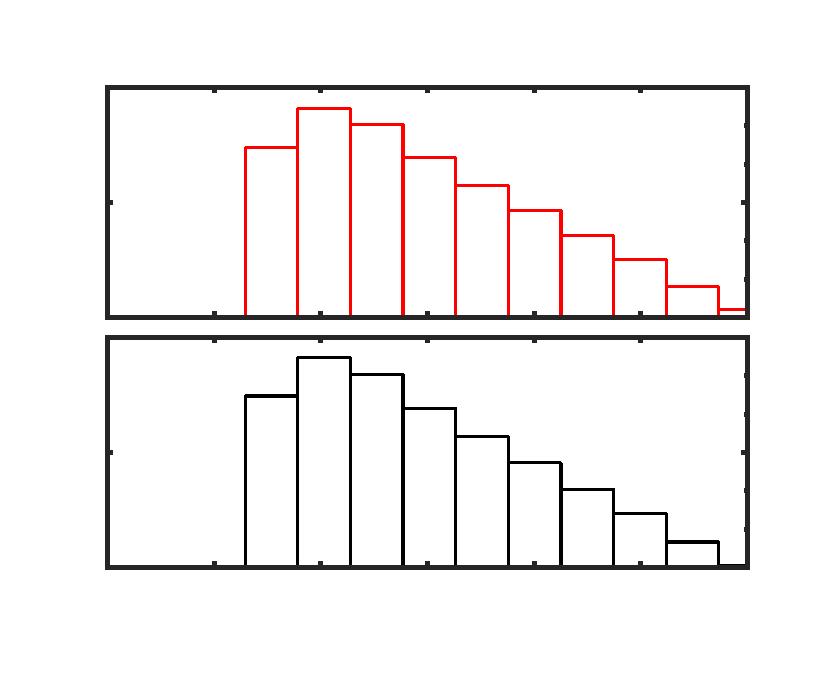}&
\includegraphics[width=1.25in]{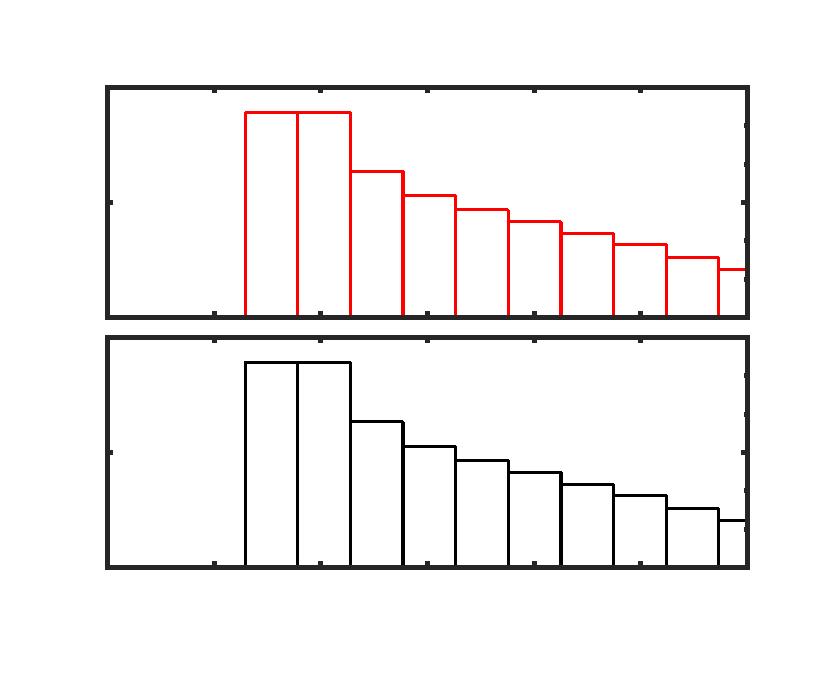}\\
\ & \textbf{(D)} Peak 3 & \textbf{(E)} Peak 4 & \textbf{(F)} Peak 5\\
\ & \includegraphics[width=1.25in]{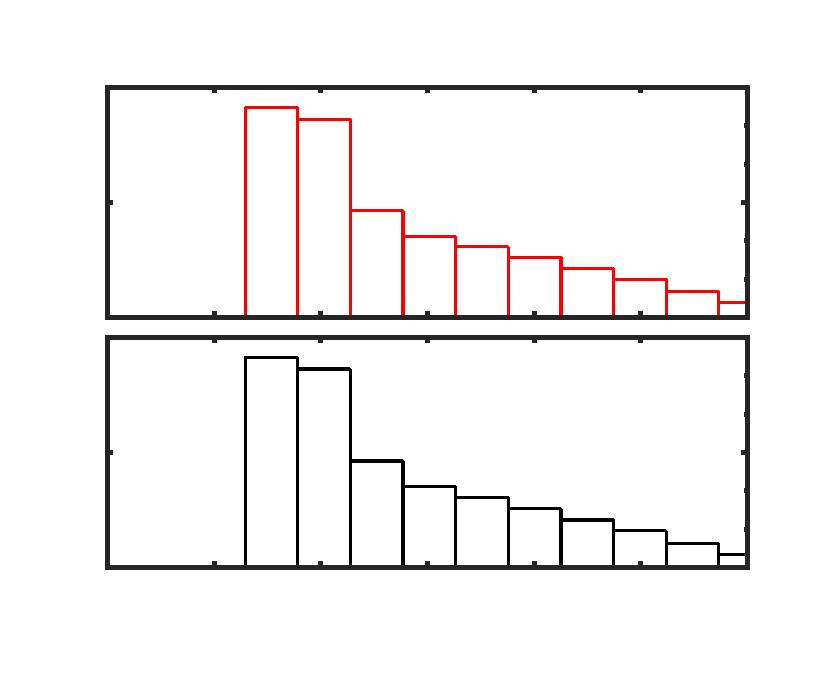}&
\includegraphics[width=1.25in]{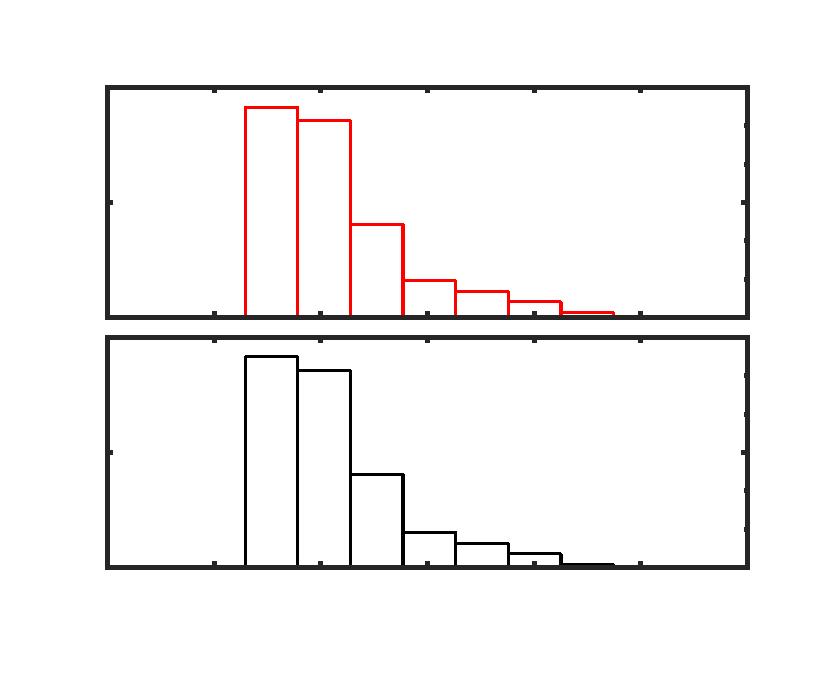}&
\includegraphics[width=1.25in]{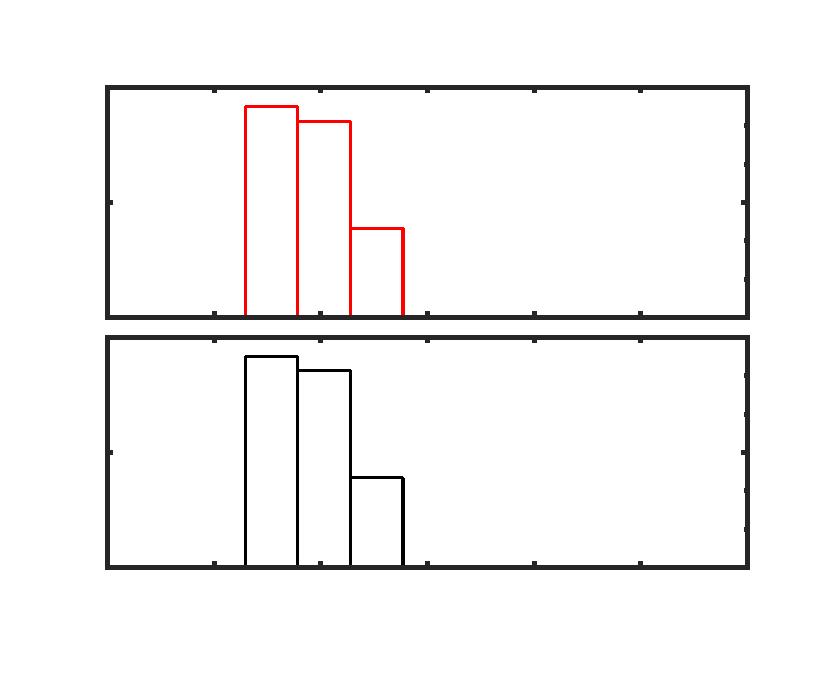}\\
\ & 
\textbf{(G)} Before pulse & \textbf{(H)} Peak 1 & \textbf{(I)} Peak 2 (missing) \\
\ &
\includegraphics[width=1.25in]{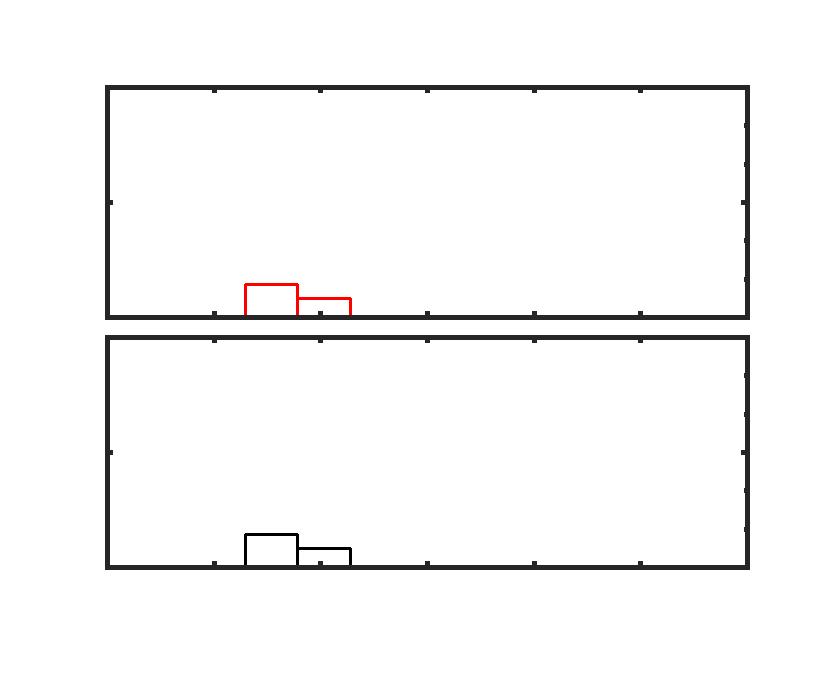}&
\includegraphics[width=1.25in]{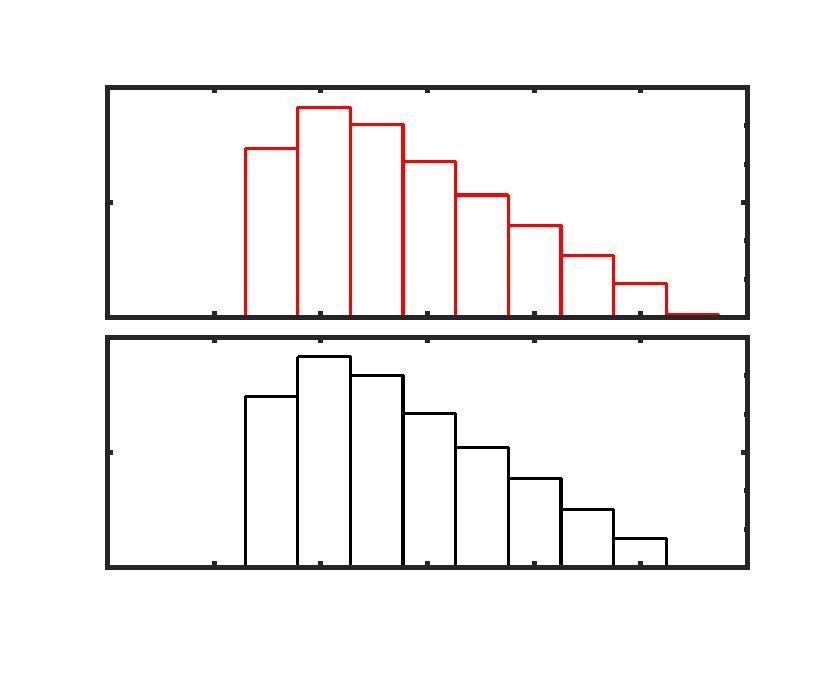}&
\includegraphics[width=1.25in]{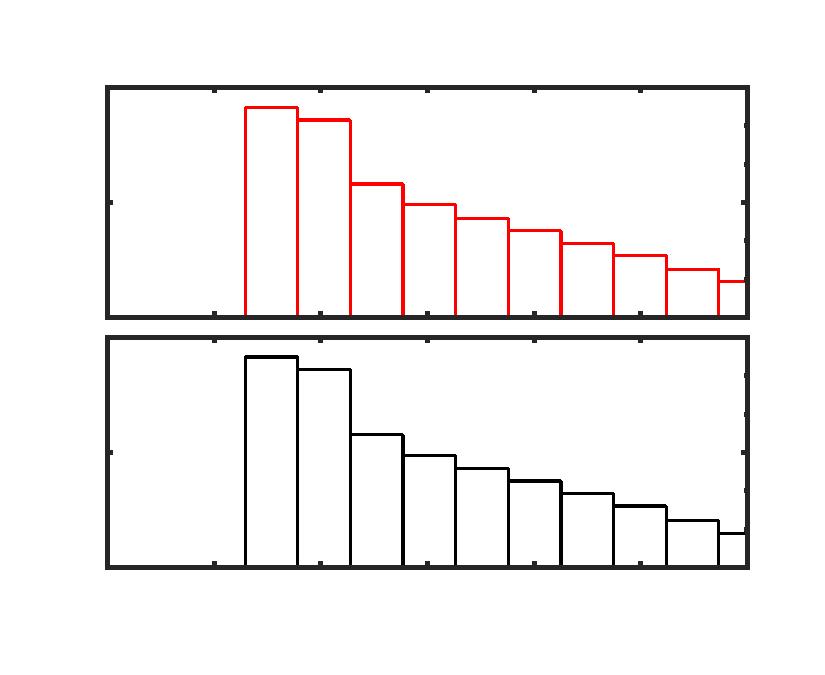}\\
\ & \textbf{(J)} Peak 3 & \textbf{(K)} Peak 4 (missing) & \textbf{(L)} Peak 5\\
\includegraphics[width=0.38in]{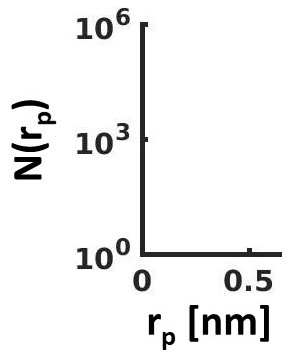}&
\includegraphics[width=1.25in]{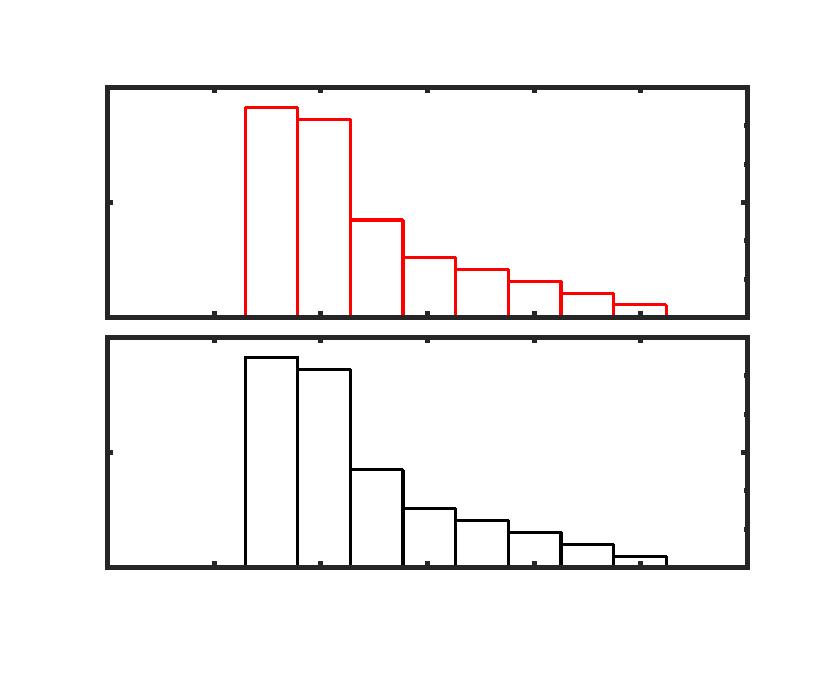}&
\includegraphics[width=1.25in]{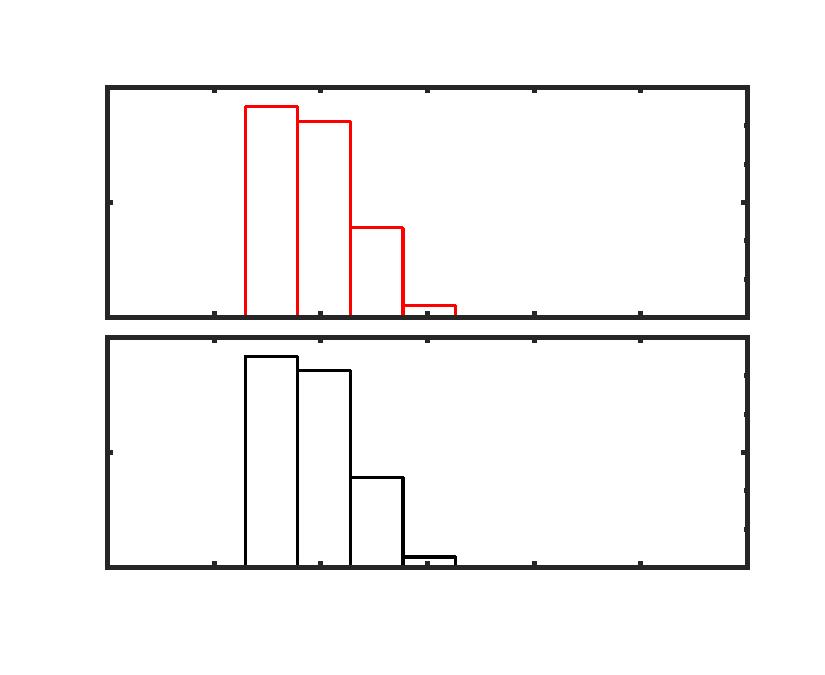}&
\includegraphics[width=1.25in]{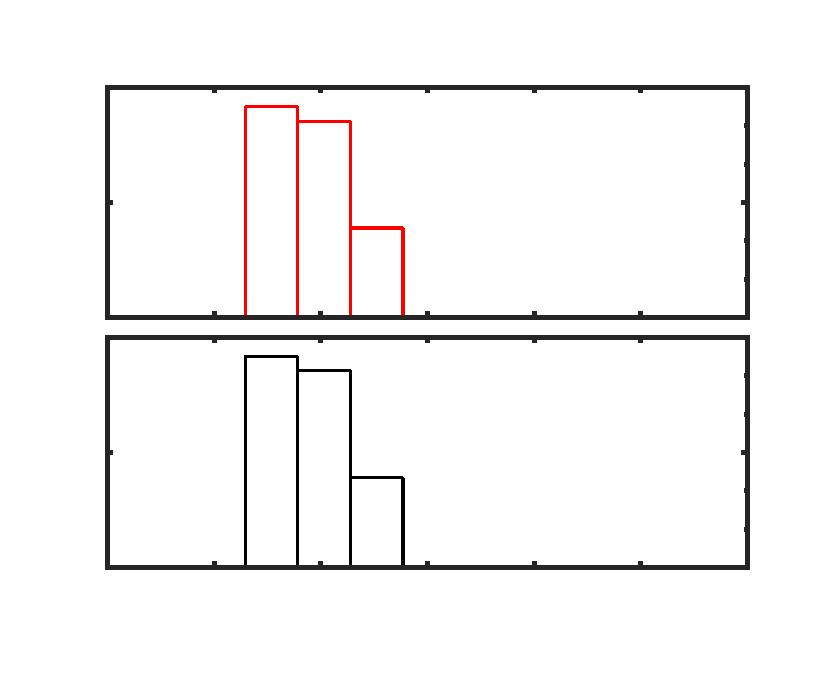}\\
\end{tabular}
\caption{\textbf{Pore histogram at the five peaks of the BP (bipolar pulse; top two rows) and the three peaks of the UP (unipolar pulse; bottom two rows).} The histograms show the number of pores as distributed by their radius (in nm).  The panels (A) and (G) show the thermalized distribution of pores at t= 0.
Pore histograms show evolution of pore distribution from the resting potentialand the associated thermalized distribution for $\mathrm{r_p\leq 3\ nm}$.  The pores are less than 2 nm in radius and the duration of the pulse is so short that very little pore expansion occurs.  These small-sized pores will provide some hindrance to the transport of Yo-Pro-1, but, any entering molecule is assumed to remain inserted in the pore.  The panels (B) and (H) occur at 180 ns, (C) and (I) at 380 ns, (D) and (J) at 580 ns, (E) and (K) at 780 ns, and (F) and (L) at 980 ns.  The times 180 ns, 380 ns, 580 ns, 780 ns and 980 ns correspond with the occurrence of the labeled peaks in Figure 2.  The pore lifetime is 100 s.  Peak 2 (I) and Peak 4 (K) of the UP are missing peaks because the UP waveform is derived from the BP waveform by rectification, as can be seen when comparing Figure 2A with Figure 2B.
}
\label{fig_Hist}
\end{figure*}


\begin{figure}[!t]
\centering
\begin{tabular}{l}
\textbf{(A)}\\ 
\includegraphics[width=4.3in]{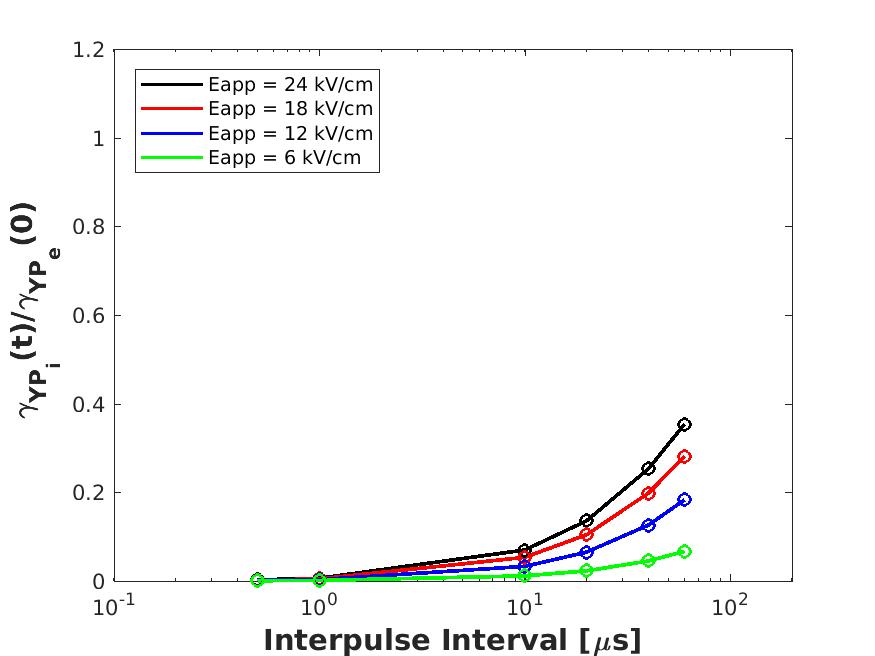}\\
\\
\textbf{(B)}\\
\includegraphics[width=4.3in]{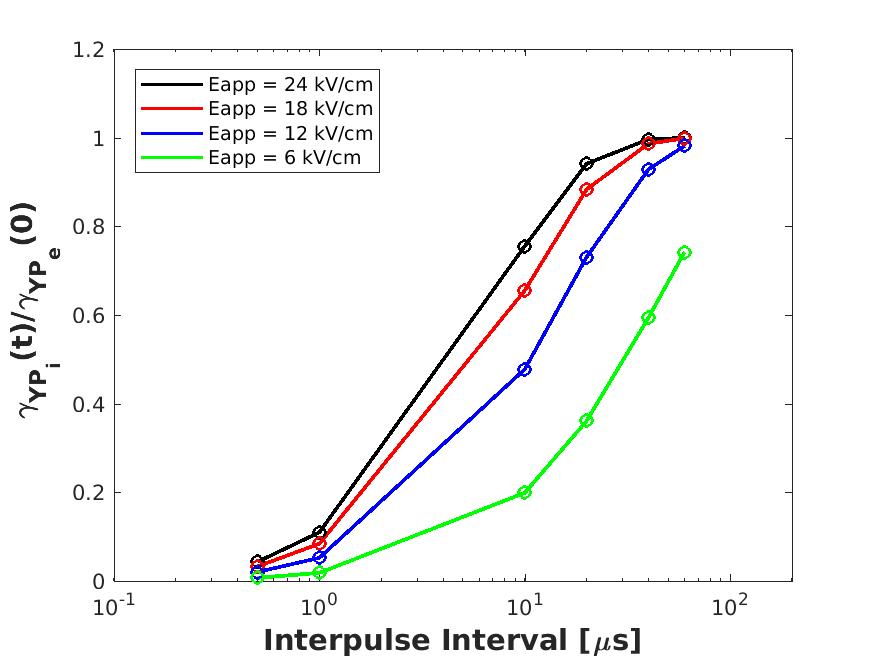}\\
\end{tabular}
\caption{\textbf{Model uptake of YP} as a function of duration between the two pulses of a bipolar pulse at t = 5 s (left) and t = 300 s(right).  The recovery of BPC effects is faster for larger field strengths.}  
\label{fig_uptake_gap_Eapp}
\end{figure}

\end{document}